

\input phyzzx.tex


\def\mod{{\hbox{\rm mod}}}

\def\vev{vacuum expectation value}

\tolerance=500000
\overfullrule=0pt
\def\np{Nucl. Phys.}
\def\pl{Phys. Lett.}

\def\cmp{Comm. Math. Phys.}

\def\mpl{Mod. Phys. Lett.}

\tolerance=500000
\overfullrule=0pt

\def\zb{{\bar z}}

\tolerance=500000
\overfullrule=0pt

\pubnum={US-FT/7-94\cr hep-th/9405151}
\date={May, 1994}
\pubtype={}
\titlepage
\title{TYPE B TOPOLOGICAL MATTER, KODAIRA-SPENCER THEORY, AND MIRROR
SYMMETRY}
\author{J.M.F. Labastida\foot{E-mail: LABASTIDA@GAES.USC.ES} and  M.
Mari\~no}
\address{Departamento de F\'\i sica de Part\'\i culas\break Universidade de
Santiago\break E-15706 Santiago de Compostela, Spain}
\abstract{Perturbing usual type B topological matter with vector
$(0,1)$-forms we find a topological theory which contains explicitly
Kodaira-Spencer deformation theory. It is shown that, in genus zero,
three-point correlation functions give the Yukawa couplings for a generic
point in the moduli space of complex structures. This generalization of
type B topological matter seems to be the correct framework to understand
mirror symmetry in terms of two-dimensional topological field theories.}

\endpage
\pagenumber=1
\sequentialequations

Topological matter in two dimensions
\REF\wtsm{E. Witten\journal\cmp&118(88)411}
\REF\ey{T. Eguchi and S.K. Yang\journal\mpl&A5(90)1693} \REF\vafa{C.
Vafa\journal\mpl&A6(91)337}
\REF\lll{J.M.F. Labastida and P.M. Llatas\journal\np&B379(92)220}
[\wtsm,\ey,\vafa,\lll] has become a very interesting framework
\REF\witmm{E. Witten, ``Mirror Manifolds and Topological Field Theory", in
{\it Essays on Mirror Manifolds}, ed. S.-T. Yau (International Press,
1992)}
\REF\vafados{C. Vafa, ``Topological Mirrors and Quantum Rings", in {\it
Essays on Mirror Manifolds}, ed. S.-T. Yau (International Press, 1992)}
[\witmm,\vafados] to understand mirror symmetry
\REF\mirror{L.J. Dixon, in {\it Superstrings, Unified Theories and
Cosmology 1987}, (G. Furlan et al., eds.) World Scientific, 1988, pag. 67}
\REF\mirdos{W. Lerche, C. Vafa and N.P. Warner\journal\np&B324(89)427}
\REF\mirtres{B.R. Greene and M.R. Plesser\journal\np&B338(90)15}
\REF\mircua{ P. Candelas, M. Lynker and R.
Schmmrigk\journal\np&B341(90)383}
\REF\alr{P.S. Aspinwall, C.A. Lutken and G.G. Ross\journal\pl&B241(90)373}
\REF\cande{P. Candelas, X.C. de la Ossa, P.S. Green and L.
Parkes\journal\np&B359(91)21 \journal\pl&B258(91)118}
\REF\mirsiete{S. Ferrara, M. Bodner and A.C. Cadavid\journal\pl&B247(90)25}
[\mirror-\mirsiete] in Calabi-Yau manifolds.  Type A topological models
are topological sigma models  whose observables depend on the moduli space
of K\"ahler forms on the target space but are independent of the complex
structure
\REF\ratio{P.S. Aspinwall and D.R. Morrison \journal\cmp&151(93)245}
[\witmm,\ratio]. Type B topological models are also topological sigma
models but in this case the observables depend on the complex structure
and are independent of the K\"ahler form [\witmm]. Mirror symmetry is
realized in this context by stating that the observables of  type A models
for a manifold are related to the observables of  type B models for its
mirror manifold.  While the dependence on the moduli space of  K\"ahler
forms in type A models is well established [\witmm,\ratio], the dependence
on the moduli space of complex structures has not yet been developed for
type B models. This is because type B models are defined [\witmm,\lll]
just at one point on moduli space. The  aim of this paper is to present a
generalization of type B models whose observables depend on the moduli
space of complex structures. As it will be shown below, this work reveals a
clear parallelism between the structures of type A and type B topological
matter.

Before describing the construction of the new form of type B topological
matter let us briefly review some results on the theory of deformations of
complex structures that will be needed in what follows. \REF\kk {K.
Kodaira, {\it Complex manifolds and deformation of complex structures},
Springer, 1985}  A standard reference on this topic is [\kk]. A complex
manifold
$M$ can be considered  as a set of  domains $\{ {\cal U}_j \}$ in ${\bf
C}^n$ glued by coordinate transformations $f_{jk}(z_k)$  which are
biholomorphic functions. A deformation of the complex structure of $M$ is
a variation of these transition functions depending on some complex
parameters $t_1,\cdots, t_s$. This deformation generates a family of
complex manifolds that will be denoted by $M_t$. It follows from this
definition that infinitesimal deformations of $M$ are always elements in
the sheaf cohomology group
$H^1(M,TM)$. However, the converse is not true. In other words,
topological obstructions to integrate an infinitesimal deformation are
found in general, and therefore  not every ${\bar {\partial}}$-closed
vector
$(0,1)$-form is associated to a family  of complex manifolds.
Alternatively, a deformation of the complex structure of $M$  can be
represented by a  vector $(0,1)$-form
$\phi (t)$ verifying the Kodaira-Spencer equation:
$$ {\bar {\partial}}\phi (t)={1 \over 2}[\phi (t),\phi (t)],
\eqn\ks
$$ and the initial condition $\phi (0)=0$. In \ks\ the bracket between a
vector
$(0,p)$-form
$\phi=\sum_{\alpha} \phi^{\alpha}\partial_{\alpha}$ and a vector
$(0,q)$-form
$\psi=\sum_{\alpha}
 \psi^{\alpha}\partial_{\alpha}$ is defined by:
$$ [\phi, \psi]=\sum_{\alpha, \beta} \Big( \phi^{\alpha} \wedge
\partial_{\alpha}
\psi^{\beta}- (-1)^{pq} \psi^{\alpha} \wedge \partial_{\alpha} \phi^{\beta}
\Big) \partial_{\beta}.
\eqn\bra
$$ In this framework, the problem of topological obstructions can be
formulated as follows: one can show that every $\phi(t)$ representing a
complex deformation verifies
 ${({\partial \phi (t) \over \partial t})}_{t=0} \in H^1(M,TM)$,
 so an infinitesimal deformation $\phi_1$ is  unobstructed if one can find
a solution $\phi(t)$ to \ks\ such that that
 ${({\partial \phi(t) \over \partial t})}_{t=0}=\phi_1$. It has been proven
by Tian
\REF\ti{G. Tian, in  S.-T. Yau (ed.), {\it Mathematical Aspects of String
Theory}, World Scientific,1987; G. Tian, in  S.-T. Yau (ed.), {\it Essays
on Mirror Manifolds}, International Press, 1992}
\REF\tod{A.N. Todorov \journal\cmp&126(89)325} [\ti] and Todorov [\tod]
that, when $M$ is a Calabi-Yau manifold, every infinitesimal  deformation
is unobstructed. More precisely, given
$\phi_1 \in H^1(M,TM)$, there is a power expansion in a parameter $t$
(which can  be understood as a parameter in the moduli space of complex
structures):
$$
\phi(t)=\phi_1 t +\phi_2t^2+\cdots
\eqn\series
$$ such that $\phi(t)$ satisfies \ks\ and therefore corresponds to a
deformation of the complex structure. In fact, the vector $(0,1)$-forms
appearing in this series are obtained by solving inductively the
Kodaira-Spencer equation at order $n$ in $t$:
$$ {\bar \partial}\phi_n={1 \over 2}\sum_{i=1}^{n-1} [\phi_i, \phi_{n-i}].
\eqn\ksn
$$ In this way, when $M$ is a Calabi-Yau, deformations of the complex
structure are in one  to one correspondence with ${\bar {\partial}}$-closed
vector $(0,1)$-forms.

In this paper we will  construct a generalization of type B topological
matter  which explicitly contains Kodaira-Spencer deformation theory and,
in some way, also Tian's results. Let us first present a brief account of
the original model. There are two equivalent ways two obtain the two
types, A and B, of topological matter by twisting $N=2$ supersymmetry. One
consists of taking an $N=2$ chiral multiplet and twist with each of the
two $U(1)$ chiral currents [\witmm]. The other one consists of twisting by
only one of the $U(1)$ chiral currents both, the $N=2$ chiral multiplet
and the $N=2$ twisted chiral multiplet [\lll]. In either case we will be
considering the B model. The target space $M$ must be a Calabi-Yau
manifold in order to avoid anomalies [\witmm]. The field content of this
model consists of a set of local  coordinates, $x^I$ and $x^{\bar I}$,
which describe a map from a Riemann surface $\Sigma$ to $M$, anticommuting
fields
$\eta^{\bar I}$, $\theta_I$ and  $\rho_\mu^I$, and auxiliary fields
$F^I$ and $F^{\bar I}$. We follow mainly the notation in [\lll] after
making the field redefinition of the original fields $\chi^{\bar I}$ and
$\bar\chi^{\bar I}$ suggested in [\witmm]: $$
\eqalign{
\eta^{\bar I}&=\chi^{\bar I}+\bar {\chi}^{\bar I}, \cr
\theta_I &=g_{I \bar J}(\chi^{\bar J}-\bar {\chi}^{\bar J}), \cr}
\eqn\redef
$$ being $g_{I \bar J}$ a metric on $M$. The model possesses the following
topological $Q$-symmetry:
$$
\eqalign{ [Q,x^I] & = 0, \cr [Q,x^{\bar I}] & = \eta^{\bar I}, \cr
\{Q,\rho_z^I\} & = \partial_z x^I,\cr
\{Q,\rho_\zb^I\} & = \partial_\zb x^I,\cr}
\qquad\qquad
\eqalign{
\{Q,\eta^{\bar I}\} & = 0,\cr
\{Q,\theta_{I}\} & = g_{I \bar J} F^{\bar J}, \cr [Q,F^I] & =
D_z\rho_\zb^I - D_\zb\rho_z^I + R^I{}_{J \bar L K}
\eta^{\bar L}
\rho_z^{J} \rho_\zb ^K, \cr [Q,F^{\bar I}] & = -\Gamma_{\bar J\bar
K}^{\bar I} \eta^{\bar J} F^{\bar K}, \cr}
\eqn\transq
$$ whose operator satisfies $Q^2=0$. The action of the type B model takes
the form:
$$
\eqalign{ S=\int_\Sigma d^2z \Big[& g_{I \bar J}\big(\partial_z x^I
\partial_\zb x^{\bar J}+ \partial_\zb x^I \partial_z x^{\bar J} \big) -
\rho_z^I \big( g_{I \bar J}D_\zb \eta^{\bar J} + D_\zb \theta_I \big) \cr
&
\,\,\,\,\,\,\,\, - \rho_\zb^I \big( g_{I \bar J}D_z
\eta^{\bar J}- D_z \theta_I \big)- R^I{}_{J \bar L K}\eta^{\bar L}
\rho_z^{J} \rho_\zb ^K \theta_I -g_{I \bar J}F^I F^{\bar J} \Big], \cr}
\eqn\action
$$ which is easily seen to be $Q$-exact:
$$  S=\Big\{ Q, \int_\Sigma d^2z \big[ g_{I \bar J} \big( \rho_z^I
\partial_\zb x^{\bar
 J}+\rho_\zb^I \partial_z x^{\bar J} \big) - F^I \theta_I \big] \Big\}.
\eqn\exacta
$$

This last relation implies that the model under consideration possesses two
of the most common features of topological quantum field theories. On one
hand, the energy-momentum tensor is  $Q$-exact. On the other hand,  the
partition function and the vacuum expectation value of products of
$Q$-invariant operators  can be  evaluated in the weak coupling regime.
This last observation implies that the observables of the theory do not
depend on the K\"ahler class chosen on $M$. Notice however that, as
pointed out in [\witmm], the type B model depends on the complex structure
of the target manifold $M$ through the
$Q$-transformation, which depends crucially on the assignation of
holomorphic and antiholomorphic coordinates. The model is defined only for
a particular point in the moduli space of complex structures,
corresponding to the Calabi-Yau manifold $M$ which we have chosen as base
point. The situation in type A topological matter is quite different:
there the action can be written as a $Q$-exact term plus a topological
term, involving only the pullback of the K\"ahler class. This remaining
piece allows one to include a explicit parametrization of the moduli space
of K\"ahler forms (as one can see, for example, in  [\ratio]). The
$Q$-symmetry for type A models does not depend on the complex structure
and therefore observables are functions on the moduli space of K\"ahler
forms. It is difficult to exploit mirror symmetry by means of topological
matter using as the partner of the type A model the usual type B model,
because moduli space structures seems to be essential to mirror symmetry
as well as to mirror computations \REF\mod{P. Candelas and X. de la Ossa
\journal\np&B355(91)455}
 [\mod,\cande,\alr]. As suggested in [\witmm], mirror symmetry should be
better understood in the framework of topological field theory if one
considers not only the  original lagrangian, but the topological family
obtained adding perturbations to it. In this way, the generic lagrangian
corresponding to the topological family should depend on a set of
parameters related, if we take the appropriate perturbation, to the moduli
space of complex structures.

The standard procedure for perturbing topological theories is the use of
descent equations [\wtsm, \witmm, \lll]. As advocated in [\lll], a very
useful tool in generating the fields entering descent equations is the
vector operator $G_\mu$. $N=2$ supersymmetry in two dimensions possesses
four spinor generators. After the twisting, two of these generators lead
to two scalar operators, being $Q$ one of them, and a vector operator
$G_\mu$. For twisted models to have the symmetry generated by
$G_\mu$ the existence of covariantly constant vectors on $\Sigma$ is
required. This requirement does not hold in general and therefore typically
the $G_\mu$ symmetry is disregarded. However, this symmetry is useful for
two reasons. First, as shown in [\lll], its gauging leads to the coupling
of topological matter to topological gravity. Second, it generates
operators which satisfy the descent equations. It is in this last respect
that it will be used in this work.

To understand why the operator $G_\mu$ generates solutions to the descent
equation let us recall that the topological algebra [\lll] identifies it as
the $Q$-partner of the momentum operator:
$$
\{Q,G_\mu\} = P_\mu.
\eqn\partners
$$ Starting with a field
$\phi^{(0)}(x)$ which satisfies,
$$  [Q,\phi^{(0)}(x)]=0,
\eqn\veinte
$$  one can construct other fields using the operators $G_{\mu}$. These
fields, which we will call partners or descendants, are antisymmetric
tensors defined as,
$$
\eqalign{
\phi^{(1)}_{\mu}(x)=[G_{\mu},\phi^{(0)}(x)\},\cr
\phi^{(2)}_{\mu \nu}(x)=[G_{\mu},[G_{\nu},\phi(x)\}\}.\cr}
\eqn\veintep
$$ Relation \partners\ guarantees that these operators satisfy the
topological descent equations,
$$  d \phi^{(n)} =  [Q,\phi^{(n+1)}\},
\eqn\vseis
$$  which in turn imply that  quantities like,
$$  W^{\Sigma}_\phi = \int_{\Sigma} \phi^{(2)},
\eqn\vsiete
$$  are $Q$-invariant. Operators of the form \vsiete\ are very useful
because they can be added to the lagrangian to build families of
topological models. This is the procedure that will be used to construct
the perturbed B model.

The $G_\mu$-transformations of the fields can be read from [\lll]. They
take the form:
$$
\eqalign{ [G_z,x^I] & = \rho_z^I, \cr  [G_\zb,x^I] & = \rho_\zb^I,\cr
[G_z,x^{\bar I}] & = 0, \cr [G_\zb,x^{\bar I}] & = 0,\cr
\{G_z,\rho_z^I\} & = 0,\cr
\{G_\zb,\rho_z^I\} & = -F^I+ \Gamma_{ J K}^{ I}\rho_z^J \rho_\zb^K,\cr
\{G_z,\rho_\zb^I\} & = F^I-\Gamma_{ J K}^{ I}\rho_z^J \rho_\zb^K ,\cr
\{G_\zb,\rho_\zb^I\} & =0,\cr}
\qquad
\eqalign{
\{G_z,\eta^{\bar I}\} & = \partial_z x^{\bar I},\cr
\{G_\zb,\eta^{\bar I}\} & =\partial_\zb x^{\bar I} ,\cr
\{G_z,\theta_{I}\} & =g_{I \bar J}\partial_z x^{\bar J}+\Gamma_{ I K}^{
J}\rho_z^K \theta_{J},\cr
 \{G_\zb,\theta_{I}\} & = g_{I \bar J}\partial_\zb x^{\bar J}+\Gamma_{ I
K}^{ J}\rho_\zb^K \theta_{J},\cr [G_z,F^I] & = -\Gamma_{ J K}^{ I}\rho_z^J
F^K, \cr [G_\zb,F^I] & =\Gamma_{ J K}^{ I}\rho_\zb^J F^K, \cr [G_z,F^{\bar
I}] & = -D_z(\eta^{\bar I}-g^{\bar I J}\theta_J)+R^{\bar I J}{}_{L
\bar K} \rho_z^L \eta^{\bar K}\theta_J, \cr [G_\zb,F^{\bar I}] &
=D_z(\eta^{\bar I}+g^{\bar I J}\theta_J)+R^{\bar I J}{}_{L \bar K}
\rho_\zb^L
\eta^{\bar K}\theta_J. \cr}
\eqn\bdiez
$$

The observables of type B topological matter have the form:
$$
\phi^{(0)}=A_{{\bar I}_1 \cdots {\bar I}_p}{} ^{J_1 \cdots J_q}\eta^{{\bar
I}_1} \cdots
\eta^{{\bar I}_p}\theta_{J_1} \cdots\theta_{J_q},
\eqn\obser
$$  where $A$ is an element in $H^p(M,\wedge^q TM)$. Notice that these are
observables only on-shell, \ie, taking into account the field equations,
because the action of $Q$ on $\theta_I$ gives terms proportional to
$\delta S
 \over \delta F^J$. As we mentioned above, Kodaira-Spencer theory and
Tian's results imply that closed vector $(0,1)$-forms are in one to one
correspondence with deformations of the complex structure, so we expect
that the appropriate topological family will be obtained perturbing with
observables of the form:
$$
\phi^{(0)}=A_{\bar I}{}^{J}\eta^{\bar I} \theta_J,
\eqn\obs
$$ where $\partial_{[{\bar K} }A_{{\bar I}]}{}^{J} =0$. In this case one
finds:
$$  [Q,\phi^{(0)}]=A_{\bar I}{}^{J}\eta^{\bar I}{\delta S \over \delta
F^J}.
\eqn\perez
$$ The fact that $\phi^{(0)}$ is $Q$-closed only on-shell has important
consequences [\witmm]: now the topological descent equations are true only
modulo terms proportional to the field equations, and this implies in turn
that we must change the topological operator $Q$ if we want the perturbed
lagrangian to be $Q$-closed. To obtain the form of the descent equations,
including the additional terms, we will use the operator $G_\mu$ introduced
above. Taking into account \partners, one finds that the operators
$\phi^{(1)}_{\mu}$ and $\phi^{(2)}_{\mu\nu}$ defined in terms of an
operator
$\phi^{(0)}$ as in \veintep\ satisfy the following modified descent
equations:
$$
\eqalign{
[Q,\phi^{(1)}_{\mu}\}&=[P_\mu,\phi^{(0)}]+[G_{\mu},[Q,\phi^{(0)}\}\},\cr
[Q,\phi^{(2)}_{\mu
\nu}\}&=[P_\mu,\phi^{(1)}_{\nu}]-[P_{\nu},\phi^{(1)}_{\mu}]+
[G_{\mu},[G_{\nu},[Q,\phi^{(0)}\} \}\}.\cr}
\eqn\desdos
$$  For an operator like \obs\ satisfying \perez\ the last terms of these
equations represent additional contributions to the standard topological
descent equations \vseis. These extra terms are proportional to field
equations due to \perez\ and the fact that the action \action\ is $G_\mu$
invariant. Actually, the simplest way to compute the form of these extra
terms is to us this invariance of the action \action. Let us denote
generically by $\psi_i$ the fields in the theory, and by $\delta S$ the
variation of the action due to an arbitrary variation $\delta
\psi_i$ of the fields. The invariance of the action \action\ under $G_\mu$
implies:
$$ [G_{\mu},\delta S]=[G_{\mu},\sum_i {\delta S \over \delta \psi_i}
\}\delta
\psi_i+ \sum_i{\delta S \over \delta \psi_i} [G_{\mu},\delta \psi_i \}=0.
\eqn\truco
$$  The useful consequence of this relation is that the transformation
under
$G_\mu$ of the field equations can be read easily after using \bdiez.

To build the perturbation of the action \action\ notice that since the
second term of the second equation in \desdos\ is linear in the field
equations, \ie,
$$  [Q,\phi^{(2)} \}=d\phi^{(1)}+\sum_i{\delta S \over \delta
\psi_i}\zeta_i,
\eqn\destres
$$  where $\zeta_i$ are quantities to be determined,  the generalized
action,
$$  S(t)=S+t\int_{\Sigma}\phi^{(2)},
\eqn\genaction
$$  is invariant up to terms of order $t^2$ under the new topological
symmetry,
$$
 [Q_t,\psi_i]=[Q,\psi_i] -t\zeta_i.
\eqn\qmod
$$

Let us study the perturbation associated to a closed vector $(0,1)$-form
as in \obs. Using \veintep\ one finds that the descendants are:
$$
\eqalign{
\phi^{(1)}_z &= D_{K}A_{\bar I}{}^{J}\rho^{K}_z \eta^{I}\theta_J +A_{\bar
I}{}^{J} \big( \partial_z x^{\bar I} \theta_J +g_{J \bar K}
\partial_z x^{\bar K}\eta^{\bar I}  \big),\cr
\phi^{(2)}_{z \zb} &=D_{L}D_{K}A_{\bar I}{}^{J}\eta^{\bar I}\rho_z^{L}
\rho_\zb ^K \theta_J +D_{K}A_{\bar I}{}^{J} \big(\rho_z^K\partial_\zb
x^{\bar I} -\rho^K_\zb \partial_z x^{\bar I} \big) \theta_J \cr &+
D_{K}A_{\bar I}{}^{J}
\big(\rho_z^K\partial_\zb x^{\bar L} -\rho^K_\zb \partial_z x^{\bar L}
\big) g_{J \bar L} \eta^{\bar I}+ g_{J \bar K}A_{\bar I}{}^{J} \big(
\partial_z x^{\bar I} \partial_\zb x^{\bar K}+ \partial_\zb x^{\bar I}
\partial_z x^{\bar K} \big)\cr &+ D_{K}A_{\bar I}{}^{J}F^K\eta^{\bar
I}\theta_J.\cr}
\eqn\desc
$$  To obtain the quantities $\zeta_i$ in \destres\ we use \truco\ for the
field equations needed. Of course, this could be obtained acting directly
with $G_\mu$ on the field equations. However, the use of \truco\ simplifies
notably the computations.  One obtains,
$$
\eqalign{
\{G_\zb, {\delta S \over \delta F^I}\}&={\delta S \over \delta \rho^I_z} -
\Gamma_{ I J}^{K}\rho_{\zb}^J{\delta S \over \delta F^K},\cr
\{G_z, {\delta S \over \delta F^I}\}&=-{\delta S \over \delta \rho^I_\zb} +
\Gamma_{ I J}^{K}\rho_z^J{\delta S \over \delta F^K},\cr
\{G_z,{\delta S \over \delta \rho^I_z} \}&={\delta S \over \delta x^I}
 -\Gamma_{ I K}^{J}\rho_{\zb}^K{\delta S \over \delta \rho_\zb^J}
+\Gamma_{ I J}^{K}\theta_K{\delta S \over \delta \theta_J} -\Gamma_{ I
K}^{J}F^K{\delta S
\over \delta F^J}  +R^{\bar L J}{}_{I \bar K} \eta^{\bar K} \theta_J
{\delta S \over \delta F^{\bar L}},\cr}
\eqn\retruco
$$     and, finally, the quantities $\zeta_i$ are read off from \destres:
$$
\eqalign{
\zeta_{x^J}=&A_{\bar I}{}^{J}\eta^{\bar I},\cr
\zeta_{\theta_J}=&-A_{\bar I}{}^{K}\eta^{\bar I}\Gamma_{J K}^{L}\theta_L,
\cr
\zeta_{\rho_z^J}=&-\partial_K A_{\bar I}{}^{J}\rho_z^K \eta^{\bar I}
-A_{\bar I}{}^{J} \partial_z x^{\bar I},\cr
\zeta_{\rho_\zb^J}=&-\partial_K A_{\bar I}{}^{J}\rho_\zb^K \eta^{\bar I}
-A_{\bar I}{}^{J} \partial_\zb x^{\bar I},\cr
\zeta_{F^J}=&D_{L}D_{K}A_{\bar I}{}^{J}\eta^{\bar I}\rho_z^{L} \rho_\zb ^K
+D_{K}A_{\bar I}{}^{J} \big(\rho_z^K\partial_\zb x^{\bar I} -\rho^K_\zb
\partial_z x^{\bar I} \big) +\partial_{K}A_{\bar I}{}^{J}F^K\eta^{\bar
I},\cr
\zeta_{F^{\bar J}}=&A_{\bar I}{}^{M}R^{\bar J L}{}_{M \bar K}\eta^{\bar I}
\eta^{\bar K}\theta_L,\cr}
\eqn\lola
$$  while all the rest vanish.

As  discussed above, adding $-t\zeta_i$ to the old $Q$ one obtains a new
one which only works at order $t$. It is natural to ask under what
conditions we can construct a topological symmetry $Q'$ verifying $Q'^2=0$
at every order in $t$. The previous  construction suggests to take:
$$
\eqalign{ [Q',x^J] & = -A_{\bar I}{}^{J}\eta^{\bar I}, \cr  [Q',x^{\bar
J}] & = \eta^{\bar J}, \cr
\{Q',\rho_z^J\} & = \partial_z x^J+\partial_K A_{\bar I}{}^{J}\rho_z^K
\eta^{\bar I} +A_{\bar I}{}^{J} \partial_z x^{\bar I}=\partial_z
x^J+\{G_z, A_{\bar I}{}^{J}\eta^{\bar I} \},\cr
\{Q',\rho_\zb^J\} & = \partial_\zb x^J+\partial_K A_{\bar
I}{}^{J}\rho_\zb^K
\eta^{\bar I} +A_{\bar I}{}^{J} \partial_\zb x^{\bar I}=\partial_\zb
x^J+\{G_\zb, A_{\bar I}{}^{J}\eta^{\bar I} \},\cr
\{Q',\eta^{\bar J}\} & = 0,\cr
\{Q',\theta_{I}\} & = g_{I \bar J} F^{\bar J}+
\partial_{I}A_{\bar K}{}^{J}\eta^{\bar K}\theta_J,\cr  [Q',F^I] & =
D_z\rho_\zb^I - D_\zb\rho_z^I + R^I{}_{J \bar L K} \eta^{\bar L}
\rho_z^{J} \rho_\zb ^K - D_{L}D_{K}A_{\bar J}{}^{I}\eta^{\bar J}\rho_z^{L}
\rho_\zb ^K \cr & -D_{K}A_{\bar J}{}^{I}
\big(\rho_z^K\partial_\zb x^{\bar J}
 +\rho^K_\zb \partial_z x^{\bar J} \big) -\partial_{K}A_{\bar
J}{}^{I}F^K\eta^{\bar J},\cr  [Q',F^{\bar J}] & = - F^{\bar L}\eta^{\bar
K}\big(\Gamma_{\bar L \bar K}^{\bar J} -g^{\bar J I}g_{\bar L M}D_I
A_{\bar K}{}^M \big), \cr}
\eqn\newal
$$ where the field $F^{\bar J}$ has been redefined in such a way that the
terms involving auxiliary fields in the final action reduce to the
quadratic form $g_{I \bar J}F^I F^{\bar J}$:
$$ F^{\bar J} \rightarrow F^{\bar J}+g^{\bar J I}D_I A_{\bar K}{}^J
\eta^{\bar K}\theta_J.
\eqn\firstredef
$$

Imposing $\{Q',[Q',x^J]\}=0$ gives the following constraint on
$A_{\bar I}{}^J$:
$$
\partial_{\bar K} A_{\bar I}{}^J \eta^{\bar K}\eta^{\bar I}-
 A_{\bar L}{}^K\partial_K A_{\bar I}{}^J  \eta^{\bar L}\eta^{\bar I}=0.
\eqn\ksdos
$$  This is just the Kodaira-Spencer equation \ks\ for the vector
$(0,1)$-form
$A_{\bar I}{}^J$. Surprisingly enough, this is a necessary and sufficient
condition for $Q'$ to be a nilpotent operator. To see this notice first
that the operators $G_{\mu}$ still close a topological algebra with $Q'$,
{\it i.e.},
$\{Q',G_{\mu} \}=P_{\mu}$. Using this fact it is easy to prove nilpotency
of
$Q'$ on $\rho_\mu^I$. For $F^J$, computations are in principle more
involved, but the field,
$$
\tilde F^I=F^I-\Gamma^I_{J K}\rho_z^J \rho^K_\zb,
\eqn\theredef
$$  has the simple transformation,
$$  [Q',\tilde F^I]=\partial_z\rho^I_\zb-\partial_\zb\rho^I_z-[G_z,\{G_\zb,
A_{\bar J}{}^I\eta^{\bar J}\}],
\eqn\simpletransf
$$  and, again the nilpotency of $Q'$ on $\tilde F^J$ can be easily proved
using the topological algebra. Finally, for the field $F^{\bar J}$ an
explicit computation leads also to the same conclusion.

Given a vector $(0,1)$-form which satisfies the  Kodaira-Spencer equation
we have at our disposal a nilpotent operator $Q'$ which satisfies the
topological algebra and reduce to the original one setting the vector
$(0,1)$-form $A_{\bar I}{}^J$ to zero. Having an off-shell nilpotent
operator like $Q'$, the simplest way to construct a $Q'$-invariant action
is to take one which is a
$Q'$-transformation. We will define the modified action as in
\exacta, but using now $Q'$ instead of $Q$. The new action takes the form:
$$ S'=\Big\{ Q', \int_\Sigma d^2z \big[ g_{I \bar J} \big( \rho_z^I
\partial_\zb x^{\bar J}+\rho_\zb^I \partial_z x^{\bar J} \big) - F^I
\theta_I
\big] \Big\}=S+\int_{\Sigma}\phi^{(2)},
\eqn\newact
$$  where $\phi^{(2)}$ is given as in \desc, with the only difference that
now
$A_{\bar I}{}^J$ verifies the Kodaira-Spencer equation instead of being
closed. The explicit form of the new lagrangian is:
$$
\eqalign{ {\cal{L}}=&g_{I \bar J}\big(\partial_z x^I \partial_\zb x^{\bar
J}+ \partial_\zb x^I \partial_z x^{\bar J} \big) - \rho_z^I \big( g_{I \bar
J}D_\zb \eta^{\bar J} + D_\zb \theta_I \big) \cr  &- \rho_\zb^I \big( g_{I
\bar J}D_z
\eta^{\bar J}- D_z \theta_I \big)- R^I{}_{J \bar L K}\eta^{\bar L}
\rho_z^{J} \rho_\zb ^K \theta_I -g_{I \bar J}F^I F^{\bar J}\cr
&+D_{L}D_{K}A_{\bar I}{}^{J}\eta^{\bar I}\rho_z^{L} \rho_\zb ^K \theta_J
+D_{K}A_{\bar I}{}^{J} \big(\rho_z^K\partial_\zb x^{\bar I} -\rho^K_\zb
\partial_z x^{\bar I} \big) \theta_J \cr &+ D_{K}A_{\bar I}{}^{J}
\big(\rho_z^K\partial_\zb x^{\bar L} -\rho^K_\zb \partial_z x^{\bar L}
\big) g_{J \bar L} \eta^{\bar I}+ g_{J \bar K}A_{\bar I}{}^{J} \big(
\partial_z x^{\bar I} \partial_\zb x^{\bar K}+ \partial_\zb x^{\bar I}
\partial_z x^{\bar K} \big).\cr}
\eqn\tocho
$$

At this point one can ask what is the relation of this theory to the one we
constructed previously along the suggestions given in [\witmm]. Recall
that, according to Tian's theorem, when $M$ is a Calabi-Yau manifold there
is a one to one correspondence between closed vector $(0,1)$-forms and
deformations of the complex structure: the series \series\  associates to
every $\phi_1$ verifying ${\bar {\partial}}\phi_1=0$ a solution to the
 Kodaira-Spencer equation \ks. Putting this expansion for $A_{\bar
I}{}^{J}$ in \newal\ we can  consider a $Q'$ symmetry up to a given order
in $t$. In fact, as expected,  the nilpotency of $Q'$ to a given order,
say $n$, corresponds to the Kodaira-Spencer equation at this order \ksn.
When perturbing type B topological matter with an observable of the
original theory, {\it i.e.}, with a closed vector $(0,1)$-form, we remain
at first order in Tian's expansion. Taking into account all terms, we
obtain the topological theory given by
\newal\ and \tocho. Notice that the original type B model as well as  the
model obtained by perturbing to first order can be understood as special
cases of the topological theory we have constructed. We will call this
theory, in analogy with
\REF\bcov{M. Bershadsky, S. Cecotti, H. Ooguri and C. Vafa, {\it
Kodaira-Spencer Theory of Gravity and Exact Results for Quantum String
Amplitudes}, HUTP-93/A025} [\bcov], {\it Kodaira-Spencer topological
matter}.

One could also wonder if it is possible to construct the perturbed A
models  by adding the pullback of the K\"ahler class in the same way that
Kodaira-Spencer topological matter  has been built, \ie,  by adding
descendant operators to the unperturbed action. The answer to this
question is positive. The action resulting after adding the corresponding
two-form descendant to the unperturbed type A action turns out to  differ
{}from the one in which just the K\"ahler class is added in a $Q$-exact
term. Since this
$Q$-exact term can be disregarded there exist a complete paralelism
between perturbed type A and perturbed type B models.

In the rest of the paper we will briefly analyze Kodaira-Spencer
topological matter using the standard tools of  topological field theories
\REF\llldos{J.M.F. Labastida and P.M. Llatas \journal\pl&B271(91)101}
 [\wtsm,\witmm,\llldos]. First notice that the action \newact\ is
$Q'$-exact, so we can work in the weak coupling regime as in type B
models. Path integrals will be localized in field  configurations that can
be easily found using Witten's fixed point theorem
\REF\witN{E. Witten \journal\np&B371(92)191} [\witmm,\witN]. It suffices to
look for the fixed points of the $Q'$ fermionic symmetry. According to
\newal, these correspond to,
$$
\eqalign{
\eta^{\bar I}=0,\cr dx^J+A_{\bar I}{}^J dx^{\bar I}=0, \cr}
\eqn\fix
$$  where $d$ denotes the exterior differential on $\Sigma$. Notice that
the second equation in \fix\ implies that $dA_{\bar I}{}^{J} \wedge
dx^{\bar I}=0$, which in turn leads to ${\bar{\partial}}A=0$. This
contradicts Kodaira-Spencer equation, so the only solution to the
differential system in
\fix\ is $dx^I=dx^{\bar{I}}=0$, and we recover the localization of type  B
model on constant configurations. Path integrals reduce to integrals over
the  manifold $M$ and fermionic integrals over the zero modes of
$\eta^{\bar I}$,
$\theta_I$ and $\rho_\mu^I$ so we have the same constraints on ghost
number as in the type B model [\witmm].

Kodaira-Spencer topological matter gives a very different class of
observables than ordinary type B models. This is obviously related to the
fact that it describes the moduli space of complex structures by means of
the vector $(0,1)$-form $A_{\bar I}{}^{J}$. If one takes a function
$f(x^I,x^{\bar J})$, the condition of observable now reads,
$$
\partial_{\bar K}f\eta^{\bar K} - A_{\bar K}{}^J \partial_J f \eta^{\bar
K}=0,
\eqn\conuno
$$  which is just the condition for $f$ to be a holomorphic function in
the new complex structure defined by $A_{\bar I}{}^{J}$  [\kk]. One can
easily obtain the condition for $V_{\bar I}{}^J \eta^{\bar I} \theta_J$
to be an observable,
$$ [Q',V_{\bar I}{}^J \eta^{\bar I} \theta_J]=\big(\partial_{\bar K}V_{\bar
I}{}^J- A _{\bar K}{}^M \partial_M V_{\bar I}{}^J-V_{\bar K}{}^M\partial_M
A _{\bar I}{}^J
\big)\eta^{\bar K}\eta^{\bar I}\theta_J+V_{\bar I}{}^J \eta^{\bar I}{\delta
S' \over \delta F^J}=0.
\eqn\condos
$$  Modulo field equations, this is equivalent to the following equation
for the vector $(0,1)$-form $V$:
$$  {\bar{\partial}}V=[A,V],
\eqn\newobs
$$  where the bracket is the same as the one defined in \bra.   A similar
condition is obtained for a general operator of the form \obser.

As stated in our brief review on Kodaira-Spencer theory, the vector
(0,1)-form $A_{\bar I}{}^J$ depends on a set of $s$ parameters
$t_1,\cdots, t_s$, where $s$ is the dimension of the moduli space of
complex structures. Choosing the Kodaira- Spencer-Kuranishi coordinates we
have in fact the series expansion [\kk,\tod]: $$
A(t)=\sum_{i=1}^{s}A_it_i+\cdots+\sum_{i_1+\cdots+i_s=n}A_{i_1 \cdots
i_s}t_1^{i_1} \cdots t_s^{i_s}+ \cdots
\eqn\ksk
$$  where the $A_i$ are a basis for $H^1(M,TM)$ and the $A_{i_1 \cdots
i_s}$ are vector $(0,1)$-forms that always exists thanks to Tian's lemma
[\ti]. Although  $A _{\bar I}{}^J\eta^{\bar I} \theta_J$ is not itself an
observable, the first derivatives  of the vector $(0,1)$-form with respect
to the parameters $t_i$, which we will denote by $\partial_i A$, verify
\newobs, as one can see taking the derivative of the Kodaira-Spencer
equation
\ks\ with respect to $t_i$.

Let us analyze some of the relevant vacuum expectation values when
$\Sigma$ is a genus zero Riemann surface and $M$ is a Calabi-Yau threefold.
In this case there are $\eta^{\bar I}$ and $\theta_I$ zero modes but no
$\rho_\mu^I$ zero modes. The ghost number selection rule allows non zero
three-point correlation functions for  observables of the form
$A _{\bar I}{}^J\eta^{\bar I} \theta_J$. After the integration of the
constant anticommuting zero modes the path integral reduces to an integral
over the manifold $M$. The correlation  function is obtained multiplying
the antisymmetrized product $\partial_iA \wedge\partial_jA \wedge
\partial_kA$ by the square of the holomorphic $(3,0)$-form $\Omega_0$
associated to the Calabi-Yau manifold $M$:
$$
\langle \partial_i A_{\bar I}{}^J\eta^{\bar I} \theta_J \partial_j A_{\bar
K}{}^L\eta^{\bar K} \theta_L
\partial_kA_{\bar M}{}^N\eta^{\bar M} \theta_N\rangle=
 \int_M \big[(\partial_iA \wedge\partial_jA  \wedge
\partial_kA ) \perp \Omega_0 \big] \wedge \Omega_0,
\eqn\vev
$$  where $\perp$ denotes the inner product. Notice that this prescription
for the computation of the path integral is defined up to a  numerical
normalization factor that can be understood as the   freedom in the choice
of the normalization of the $(3,0)$-form $\Omega_0$ [\witmm]. As shown in
[\bcov], this are, again up to a factor, the Yukawa  couplings
$C_{ijk}(t)$, {\it i.e.}, the Yukawa couplings evaluated at a point of the
moduli space of complex structures parametrized by $t$. This follows from
Todorov's expression  for the deformation of the holomorphic $(3,0)$-form
as one changes the complex structure [\tod], $$
\Omega_t=\Omega_0+A\perp \Omega_0-(A\wedge A)\perp \Omega_0-(A\wedge A
\wedge A)\perp \Omega_0,
\eqn\todoex
$$ and the following expression for the Yukawa couplings [\mod, \cande]:
$$ C_{ijk}(t)=-\int_M \Omega_t \wedge \partial_i \partial_j \partial_k
\Omega_t.
\eqn\yukawa
$$

Notice that the mirror computation in [\cande] involves precisely the
Yukawa couplings for a  generic point in the space of the complex
structures, and compares them with an expansion depending on the
parameters of the moduli space of K\"ahler forms. It has been shown
[\ratio] that type A topological matter naturally gives rise to an
expansion for the three-point correlation functions which have the
polynomial form:
$$  f_{ijk}=
N^0_{ijk}+N^1_{abc}q_1+N^2_{ijk}q_2+\cdots+N^{\cdots}_{ijk}q_1^2q_2+\cdots
\eqn\expA
$$  The coefficients $N^v_{ijk}$ are intersection numbers of cycles in a
moduli space of holomorphic maps from $\Sigma$ to $M$, and the $q_i$ are
parameters for the K\"ahler form. Now it seems that it is Kodaira-Spencer
topological matter  rather than the usual type B topological matter the one
which naturally gives the mirror quantity computed in  [\cande] and
contains the information about the moduli space of complex structures as
encoded in
$A$. Moreover, mirror symmetry between type A and type B models seems
enhanced when taking Kodaira-Spencer topological matter: if one
substitutes \ksk\ in
\vev, the following expansion is obtained for the Yukawa coupling:
$$
\eqalign{ &C_{ijk}(t)=\int_M \big[ ( A_i\wedge A_j \wedge A_k )\perp
\Omega_0 \big] \wedge \Omega_0\cr &+\Bigg(\int_M \big[ ( A_{1i} \wedge A_j
\wedge A_k ) \perp \Omega_0 \big] \wedge \Omega_0 + \int_M \big[ ( A_i
\wedge A_{1j}  \wedge A_k)\perp \Omega_0 \big] \wedge
\Omega_0+\cdots \Bigg)t_1\cr  &+\cdots \cr}
\eqn\expB
$$  In this expansion the coefficients are integer combinations of the
various couplings between the vector $(0,1)$-forms which appear in Tian's
expansion of the solution to the Kodaira-Spencer
 equation, and the
$t_i$ are the Kodaira-Spencer-Kuranishi parameters of the complex
structure. One can regard
\expA\ and \expB\ as mirror expansions and exploit this fact once the
mirror map relating complex and K\"ahler parameters is known.

The results presented here give rise to many issues. Of course, it is
worth pursuing the study of Kodaira-Spencer topological matter by itself or
as an intermediate step to understand the extended moduli space of
topological sigma models [\witmm]. This seems to be a promising approach to
the mirror symmetry and to the mirror map between moduli spaces from the
point of view of topological field theories. But perhaps the most urgent
question is the relation  of Kodaira-Spencer topological matter with the
Kodaira-Spencer theory of gravity presented in [\bcov]. It would be
interesting to know if the topological theory we have constructed,  which
includes the field $A_{\bar I}{}^J$ as a background field, gives rise to
the six-dimensional action of Kodaira-Spencer theory of gravity. Notice
that the third derivatives of this action with respect to the moduli
parameters give, at the tree level, the Yukawa couplings [\bcov]. In
\REF\dvv{R. Dijkgraaf, H. Verlinde and E. Verlinde \journal\np&B352(90)59}
[\dvv] it was shown in the framework of topological conformal field
theories that perturbations of a topological theory can be encoded in a
free energy function $F(t)$, so that three-point correlation functions at
genus zero are given by
$c_{ijk}(t)=\partial_i
\partial_j
\partial_k F(t)$. In this way the action of the Kodaira-Spencer theory of
gravity could be the analog of free energy for Kodaira-Spencer topological
matter. We expect to report on this in the future.

\vskip1cm

\ack We would like to thank A. V. Ramallo for very helpful discussions.
This work was supported in part by DGICYT under grant PB90-0772 and by
CICYT under grants AEN93-0729 and AEN94-0928.

\vskip0.5cm

\refout
\end